\documentstyle[epsfig,eqsecnum,aps]{revtex}
\begin{document}{\rm }

\title{Prescission neutron multiplicity and fission probability
from Langevin dynamics of nuclear fission}

\author{ Gargi Chaudhuri \thanks{Electronic address:
gargi@veccal.ernet.in} and Santanu Pal\thanks{Electronic address:
santanu@veccal.ernet.in}}

\address{Variable  Energy  Cyclotron  Centre,  1/AF Bidhan Nagar,
Kolkata 700 064, India}
\maketitle
\begin{abstract}

A  theoretical  model  of  one-body  nuclear  friction  which was
developed earlier, namely the  chaos-weighted  wall  formula,  is
applied  to  a dynamical description of compound nuclear decay in
the framework of the Langevin equation coupled  with  statistical
evaporation of light particles and photons. We have used both the
usual wall formula friction and its chaos-weighted version in the
Langevin  equation  to  calculate  the  fission  probability  and
prescission  neutron  multiplicity  for   the   compound   nuclei
$^{178}$W,  $^{188}$Pt,  $^{200}$Pb,  $^{213}$Fr, $^{224}$Th, and
$^{251}$Es. We  have  also  obtained  the  contributions  of  the
presaddle  and  postsaddle  neutrons  to  the  total  prescission
multiplicity. A detailed analysis of  our  results  leads  us  to
conclude  that  the  chaos-weighted  wall  formula  friction  can
adequately describe the fission dynamics in the presaddle region.
This friction, however, turns  out to be too weak to describe the
postsaddle dynamics properly. This  points  to  the  need  for  a
suitable  explanation  for  the  enhanced neutron emission in the
postsaddle stage of nuclear fission.

\end{abstract}



\section{INTRODUCTION}

The   emission   of   light  particles  and  photons  during  the
prescission stage of a fissioning nucleus  has  proved  to  be  a
useful  source  of  information regarding the dynamics of nuclear
fission  \cite{fro1}.  In   particular,   the   multiplicity   of
prescission  neutrons,  measured  over a wide range of excitation
energies  for  a  number  of  compound  nuclei,   has   confirmed
\cite{thoe}  that  the  fission  lifetime  of  a  hot  nucleus is
substantially longer than that determined  from  the  statistical
model of Bohr and Wheeler \cite{bw}. It is, therefore, natural to
expect  that  a  dissipative  dynamical  model  would  provide an
appropriate description of nuclear  fission  at  high  excitation
energies.  This has given rise to a renewed interest in the works
of Kramers \cite{kram} who considered  the  dynamics  of  nuclear
fission  to be similar to that of a Brownian particle floating in
a viscous heat bath. Though Fokker-Planck equation was  initially
used    to    describe    such   dissipative   fission   dynamics
\cite{wm1,wm2}, the application  of  the  Langevin  equation  was
found to be more convenient in the later works \cite{fro1,ab1}.

The  Langevin  equation  has  been used extensively in the recent
years \cite{fro1,ab1,ab2,fro2,fro3,pomor} in order to explain the
prescission  neutron  multiplicity  and  fission  probability  of
highly  excited  (typically a few tens of MeV and above) compound
nuclei  formed  in  heavy-ion  induced fusion reactions. In these
calculations, evaporation of the neutrons and photons (and  other
light  particles) is considered at each instant of time evolution
of the fission degrees of freedom.  One  of  the  most  important
inputs to such Langevin dynamical calculations is the dissipative
property   of   the  nucleus  since  it  accounts  for  both  the
dissipative and the random forces acting on the  fission  degrees
of  freedom. While the other inputs to the Langevin equation such
as the potential  and  inertia  can  be  obtained  from  standard
nuclear  models,  the  strength of the dissipative force is still
not  an  unambiguously  defined  quantity  and  is  often   fixed
empirically in order to fit the experimental data. In this paper,
we  shall  be  mainly  concerned with the choice of a dissipative
force, based on physical  arguments,  which  can  be  used  in  a
dynamical description of nuclear fission.

Fr\"obrich  {\it et al.} \cite{fro3} made a detailed study of the
fission dynamics and  prescission  particle  emission  using  the
Langevin   equation.  A  comparison  of  the  calculated  fission
probability  and  prescission  neutron  multiplicity   excitation
functions  for  a number of nuclei with the experimental data led
to a phenomenological shape-dependent nuclear  friction  in  this
work. The phenomenological friction turned out to be considerably
smaller  ($\sim  10\%$)  than the standard wall formula value for
nuclear  friction  for  near-spherical  shapes  of  the  compound
nucleus  whereas  a strong increase of this friction was found to
be necessary at large  deformations.  Similar  observations  were
also reported by other workers \cite{pomor} who obtained a better
agreement  with  the  experimental  values of prescission neutron
multiplicity by reducing  the  strength  of  the  wall  friction.
Earlier,  Nix  and Sierk \cite{nix1,nix2} also suggested in their
analysis  of  mean  fragment  kinetic  energy   data   that   the
dissipation  is  about  4 times weaker than that predicted by the
wall-and-window formula of one-body dissipation.

The wall formula for nuclear dissipation was invented long ago in
a  simple  classical picture by extending the mean field concepts
to the domain of dissipative dynamics \cite{blocki}. One  crucial
assumption  of the wall formula concerns the randomization of the
particle (nucleon) motion due to  the  successive  collisions  it
suffers  at  the  nuclear  surface.  The  derivation  of the wall
formula assumes that the particle motion is fully randomized.  It
was   early   realized   that   any   deviation  from  this  full
randomization assumption would give rise to a  reduction  in  the
strength of the wall formula friction \cite{blocki,koo}. However,
it  is  only recently that a modification of the wall formula has
been proposed in  which  the  full  randomization  assumption  is
relaxed  in  order  to  make it applicable to systems with partly
chaotic single-particle motion \cite{pal1}. In what  follows,  we
shall use the term "chaos- weighted wall formula" (CWWF) for this
modified  friction  in  order to distinguish it from the original
wall formula (WF) friction. As was shown in Ref.\cite{pal1},  the
CWWF friction coefficient ${\eta_{cwwf}}$ will be given as

\begin{eqnarray}
\label{fric}
\eta_{cwwf} = \mu \eta_{wf},
\label{cwwf}
\end{eqnarray}

\noindent  where ${\eta_{wf}}$ is the friction coefficient as was
given by the original wall formula \cite{blocki} and ${\mu}$ is a
measure  of  chaos (chaoticity) in the single-particle motion and
depends on the instantaneous shape of the nucleus. The  value  of
chaoticity  $\mu$ changes from 0 to 1 as the nucleus evolves from
a spherical shape to a highly deformed one. The CWWF friction  is
thus much smaller than the WF friction for compact nuclear shapes
while   they  become  closer  at  large  deformations.  Thus  the
suppression of the strength of wall formula friction achieved  in
the  chaos-weighted  wall  formula  suggests  that a lack of full
randomization or chaos in single-particle motion  can  provide  a
physical  explanation  for  the reduction in strength of friction
for compact nuclear shapes as required  in  the  phenomenological
friction of Ref.\cite{fro3}.

The  main  motivation  of  the  present work is to verify to what
extent the  chaos-weighted  wall  formula  can  account  for  the
experimental   prescission   neutron   multiplicity  and  fission
probability data. To this end, we shall use both the CWWF and  WF
frictions  as  input  to  the  Langevin  equation.  The  Langevin
equation will be solved by coupling it with neutron and  $\gamma$
evaporation  at  each  step  of its time evolution. Following the
work of Fr\"obrich {\it et  al.}  \cite{fro3},  we  shall  use  a
combined  dynamical  and statistical model for our calculation in
which a switching over to a statistical model description will be
made when the fission process reaches the stationary regime.  The
prescission  neutron multiplicity and fission probability will be
obtained  by  sampling  over   a   large   number   of   Langevin
trajectories.  We  shall  perform  calculations  at  a  number of
excitation energies for each of the  compound  nuclei  $^{178}$W,
$^{188}$Pt,$^{200}$Pb,$^{213}$Fr,  $^{224}$Th,  and $^{251}$Es. A
detailed  comparison  of   the   calculated   values   with   the
experimental data will be presented.

It is worthwhile  here  to  point  out  a  special feature of the
present work. We do not have  any  adjustable  parameter  in  our
entire  calculation. All the input parameters except the friction
coefficients are fixed by  standard  nuclear  models.  The  chaos
weighted  wall  friction  coefficient  is  obtained  following  a
specific  procedure  \cite{pal1}   which   explicitly   considers
particle  dynamics  in  phase  space  in  order  to calculate the
chaoticity factor  $\mu$  in  Eq.\ref{cwwf}.  There  is  no  free
parameter  in this calculation of friction. In fact, our main aim
in this paper is to calculate  observable  quantities  using  the
theoretically   predicted   friction   and   compare   them  with
experimental values in order to draw  conclusions  regarding  the
validity  of  the  theoretical  model  of nuclear friction. As it
would turn out,  our  calculation  would  not  only  confirm  the
theoretical  model  of the chaos weighted wall friction, it would
also  provide  physical justification for the empirical values of
friction used in other works \cite{fro3}.  The  present  work  is
thus expected to contribute significantly to our understanding of
the dissipative mechanism in nuclear fission.

The paper is organized as follows. The dynamical model along with
the  necessary  input  as used in the present calculation will be
given in the next section. The details of  the  calculation  will
also   be   given   here.   The  calculated  prescission  neutron
multiplicities and fission probabilities will  be  compared  with
the  experimental  values  in  Sec.III.  A summary of the results
along with the conclusions will be presented in the last section.

\section{DETAILS OF THE MODEL}
\subsection {Collective coordinates, potential and inertia}

We  have  discussed  the Langevin equation along with the various
input as used in the present calculation in a recent  publication
\cite{pal2}.  We  shall use the same definitions and notations in
the present work, a brief description of which  are  as  follows.
The  shape  parameters  $c,h$  and $\alpha$ as suggested by Brack
{\it  et  al.}  \cite{brack}  will  be  taken  as  the collective
coordinates for the fission degree of freedom. However,  we  will
simplify the calculation by considering  only  symmetric  fission
($\alpha=0$).  We  shall  further assume in the present work that
fission would proceed along the valley of the potential landscape
in $(c,h)$ coordinates though  we  shall  consider  the  Langevin
equation  in  elongation  $(c)$  coordinate  alone  in  order  to
simplify   the  computation.  Consequently,  the  one-dimensional
potential  in  the  Langevin  equation   will   be   defined   as
$V(c)=V(c,h)$  {\it  at valley}. Other quantities such as inertia
and friction will also  be  similarly  defined.  Since  our  main
concern in the present work is to distinguish between CWWF and WF
frictions which give rise to fission rates differing by more than
$100\%$,  and it has been already noted that the fission rates in
two-dimensional and one-dimensional cases differ by not more than
$15\%$ \cite{wada},  our  approximation  of  considering  fission
dynamics  in  one  dimension  can  be considered adequate for our
purpose. Moreover, we have  also  checked  that  the  prescission
neutron  multiplicity and fission probability change by less than
$5\%$ when  the  input  fission  rates  are  changed  by  $15\%$.
Therefore,  we  estimate that the uncertainty associated with our
calculation is rather small allowing us to  compare  our  results
with the experimental data.

We shall, therefore, proceed by considering $c$ and its conjugate
momentum  $p$  as  the  dynamical  variables  for fission for our
present study and the coupled Langevin equations in one dimension
will be given \cite{ab2} as

\begin{eqnarray}
\frac{dp}{dt}   &=&
-\frac{p^2}{2}   \frac{\partial}{\partial   c}\left({1  \over  m}
\right) -
   \frac{\partial F}{\partial c} - \eta \dot c + R(t), \nonumber\\
\frac{dc}{dt} &=& \frac{p}{m} .
\label{langv}
\end{eqnarray}

\noindent  The  shape-dependent  collective  inertia    and   the
friction coefficient in the above equations are  denoted  by  $m$
and $\eta$ respectively. The free energy of the system is denoted
by $F$ while $R(t)$ represents the random part of the interaction
between the fission degree of freedom and the rest of the nuclear
degrees  of  freedom considered collectively as a thermal bath in
the  present  picture.  The  collective  inertia,  $m$,  will  be
obtained by assuming  an  incompressible  irrotational  flow  and
making   the   Werner-Wheeler  approximation  \cite{davies}.  The
driving force in a thermodynamic system should  be  derived  from
its  free  energy  for which we will use the following expression
\cite{fro3} considering the nucleus  as  a  noninteracting  Fermi
gas:

\begin{equation}
F(c,T) = V(c) - a(c) T^2, \label{free}
\end{equation}

\noindent  where  $T$ is the temperature of the system and $a(c)$
is the coordinate-dependent level density parameter which will be
chosen following Ref.\cite{fro3}.

The  surface  of  a  nucleus  of  mass number $A$ with elongation
and neck coordinates, $c$ and $h$, is defined as

\begin{eqnarray}
\rho^2(z)  &=&  \left(1  -
\frac{z^2}{c_0^2}\right)(a_0c_0^2    +   b_0z^2),   \label{shape}
\end{eqnarray}
where
\begin{eqnarray} c_0 &=& cR, \nonumber\\
 R &=& 1.16 A^{1 \over 3} \nonumber
\end{eqnarray}

and

\begin{eqnarray}
a_0 &=& \frac{1}{c^3} - \frac{b_0}{5},\nonumber\\
b_0 &=& 2h + \frac{c-1}{2} ,\nonumber
\end{eqnarray}

\noindent in cylindrical coordinates. The potential energy $V(c)$
is  obtained from the finite-range liquid drop model \cite{sierk}
where we calculate  the  generalized  nuclear  energy  by  double
folding  the  uniform  density  within  the  above surface with a
Yukawa-plus-exponential potential. The Coulomb energy is obtained
by double  folding  another  Yukawa  function  with  the  density
distribution.   The  various  input  parameters  are  taken  from
Ref.\cite{sierk} where they were determined from fitting  fission
barriers  of  a wide range of nuclei. The centrifugal part of the
potential  is  calculated using the rigid body moment of inertia.
The potential is calculated over a grid of $(c,h)$ values and the
valley  of  the minimum potential located. Potential values along
this valley is used in solving the Langevin equation.

The  instantaneous random force $R(t)$ is modeled after that of a
typical Brownian  motion  and  is  assumed  to  have a stochastic
nature  with  a  Gaussian  distribution  whose  average  is  zero
\cite{ab1}.  It  is  further assumed that $R(t)$ has an extremely
short  correlation  time  implying  that  the  intrinsic  nuclear
dynamics  is  Markovian.  Consequently the strength of the random
force can be obtained from  the  fluctuation-dissipation  theorem
and  the  properties of $R(t)$ can be written as

\begin{eqnarray}
\langle    R(t)    \rangle    &=&    0    ,\nonumber\\
\langle
R(t)R(t^{\prime})\rangle   &=&  2\eta  T  \delta(t-t^{\prime}).
\label{random}
\end{eqnarray}

\subsection {Dissipation}

One-body  dissipation is usually considered to be more successful
in  describing   fission     dynamics   than  two-body  viscosity
\cite{ab1,ab2}.   We   shall,   therefore,   use   the   one-body
wall-and-window dissipation \cite{blocki}    in    the   Langevin
equation. For the one-body wall dissipation, we   shall  use  the
chaos-weighted wall formula  (Eq.\ref{cwwf})  introduced  in  the
preceding  section.  The  chaoticity  $\mu$ in Eq.\ref{cwwf} is a
measure of chaos in the single-particle motion  of  the  nucleons
within  the  nuclear volume and in the present classical picture,
this will be  given  as  the  average  fraction  of  the  nucleon
trajectories that are chaotic when the sampling is done uniformly
over  the nuclear surface. A trajectory is identified either as a
regular or as a chaotic one by considering the magnitude  of  its
Lyapunov  exponent and the nature of its variation with time. The
details of this procedure are given  in  Ref.\cite{blocki2}.  The
chaoticity  is  calculated  for  all  possible  shapes  up to the
scission configuration. A plot of  the  variation  of  chaoticity
with elongation can be found in Ref.\cite{pal2}.

In  the wall-and-window model of one-body dissipation, the window
friction is expected to be effective after a neck  is  formed  in
the nuclear system \cite{sierk2}. Further, the radius of the neck
connecting the two future fragments should be sufficiently narrow
in  order  to  enable a particle that has crossed the window from
one side to the other to remain within the other fragment  for  a
sufficiently  long  time. This is necessary to allow the particle
to undergo a sufficient number of  collisions  within  the  other
side  and  make  the  energy  transfer irreversible. It therefore
appears that the window friction should be very nominal when neck
formation just begins. Its strength should increase as  the  neck
becomes  narrower  reaching  its  classical  value  when the neck
radius becomes  much  smaller  than  the  typical  radii  of  the
fragments.  We  however  know  very little regarding the detailed
nature of such a transition.  We  shall  therefore  refrain  from
making any further  assumption  regarding  the  onset  of  window
friction.  Instead,  we  shall  define  a transition point in the
elongation coordinate $c_{win}$ beyond which the window  friction
will  be  switched  on.  We  shall  also assume that the compound
nucleus  evolves  into  a  binary  system  beyond  $c_{win}$  and
accordingly  correction  terms  for the motions of the centers of
mass of the two halves will be applied to the  wall  formula  for
$c>c_{win}$ \cite{sierk2}.

The  choice of a suitable value for the transition point requires
some  consideartion. We first note that while the window friction
makes  a  positive  contribution  to  the  total  wall-and-window
friction  for  $c>c_{win}$,  the center of mass motion correction
reduces the wall friction.  Therefore,  these  two  contributions
cancel   each  other  to  a  certain  extent.  Consequently,  the
resulting wall-and-window friction is not very sensitive  to  the
choice  of  the  transition  point. We shall further explore this
point  quantitatively  as follows. When a nucleus moves along the
fission path, a neck formation just begins at $c=1.5$.  Thus  the
transition  point  can  lie  anywhere  beyond this point upto the
scission configuratrion. We have performed a few calculations for
prescission neutron multiplicity  and  fission  probability  with
values  of  $c_{win}$  beyond  $1.5$ the calculated values are in
agreement within $5\%$. Therefore, the value of $c_{win}$ is  not
very  critical  for  our  purpose.  We  shall  choose a value for
$c_{win}$ at which the nucleus has a binary shape  and  the  neck
radius is half of the radius of either of the would-be fragments.
This  value  of  $c_{win}$ is thus half-way between its lower and
the upper limit in terms  of  the  neck  radius.  Though  such  a
consideration  to choose a value of $c_{win}$ is still arbitrary,
we have just demonstrated that it will have little  influence  on
our results.

We   shall   use  the  following  expressions  to  calculate  the
wall-and-window   friction   coefficients    ($\eta_{wf}$    will
henceforth   stand   for   the   full  wall-and-window  friction)
\cite{sierk2}:

\begin{eqnarray}
\eta_{wf}(c < c_{win})= \eta_{wall}(c < c_{win}),
\label{wf1}
\end{eqnarray}

\noindent where

\begin{eqnarray}
 \eta_{wall}(c < c_{win})=
{1 \over 2} \pi \rho_m {\bar v}
\int_{z_{min}}^{z_{max}} { \left( \frac{\partial \rho^2}{\partial c}
\right)}^2  {\left[\rho^2 + {\left({1 \over 2}\frac{\partial \rho^2}
{\partial z}\right)}^2\right]}^{-{1 \over 2}} dz,
\label{wall1}
\end{eqnarray}

\noindent and

\begin{eqnarray}
\eta_{wf}(c \ge c_{win})= \eta_{wall}(c \ge c_{win}) +
\eta_{win}(c \ge c_{win}),
\label{wf2}
\end{eqnarray}

\noindent where

\begin{eqnarray}
\eta_{wall}(c \ge c_{win})&=&{1 \over 2} \pi \rho_m {\bar v}
\left\{\int_{z_{min}}^{z_N} {\left( \frac{\partial \rho^2}
{\partial c} + \frac{\partial \rho^2}{\partial z}
\frac{\partial D_1}{\partial c}\right)}^2 {\left[\rho^2 +
{\left({1 \over 2}\frac{\partial \rho^2}{\partial z}\right)}^2\right]}^
{-{1 \over 2}}dz\right .\nonumber\\ &&+\left .\int_{z_N}^{z_{max}}
{\left( \frac{\partial \rho^2}{\partial c}+ \frac{\partial \rho^2}
{\partial z} \frac{\partial D_2}{\partial c}\right)}^2 {\left[\rho^2 +
{\left({1 \over 2} \frac{\partial \rho^2}{\partial z}\right)}^2\right]}^
{-{1 \over 2}}dz  \right\},
\label{wall2}
\end{eqnarray}

\noindent and
\begin{equation}
\eta_{win}(c \ge c_{win})
=    {1   \over   2}   \rho_m  {\bar  v}
{\left(\frac{\partial  R}{\partial  c}\right)}^2  \Delta  \sigma.
\label{win}
\end{equation}

 In  the  above  equations, $\rho_{m}$ is the mass density of the
nucleus, $\bar{v}$  is  the  average  nucleon  speed  inside  the
nucleus  and $D_{1}$, $D_{2}$ are the positions of the centers of
mass of the two parts of the fissioning system  relative  to  the
center  of  mass of the whole system. $z_{min}$ and $z_{max}$ are
the two extreme ends of the nuclear shape along the $z$ axis  and
$z_{N}$  is  the  position  of  the  neck  plane that divides the
nucleus into two parts. In the window friction coefficient,   $R$
is the distance between centers of mass of  future  fragments and
$\Delta \sigma$ is the area of the window between the  two  parts
of the system.

The  wall  friction  coefficients  given  by (Eqs.\ref{wall1} and
\ref{wall2}) are obtained \cite{blocki} under the assumption of a
fully  chaotic nucleon motion within the nuclear volume. However,
a fully chaotic motion is achieved only when the nuclear shape is
extremely irregular whereas the nucleon motion is partly  chaotic
in  varying  degrees  for  typical nuclear shapes through which a
nucleus evolves when it undergoes fission. We have already argued
in the preceding section that for such cases, the chaos  weighted
wall  friction  ($\eta_{cwwf}$) should be employed instead of the
original   wall   friction.   Accordingly,   we   shall   replace
Eqs.\ref{wall1}  and \ref{wall2} by their chaos weighted versions
and  the   chaos-weighted   wall-and-window   friction   (denoted
henceforth by $\eta_{cwwf}$) is subsequently obtained as

\begin{eqnarray}
\eta_{cwwf}(c < c_{win})= \mu (c) \eta_{wall}(c < c_{win}),
\label{cwwf1}
\end{eqnarray}

\noindent and

\begin{eqnarray}
\eta_{cwwf}(c \ge c_{win})= \mu(c) \eta_{wall}(c \ge c_{win}) +
\eta_{win}(c \ge c_{win}).
\label{cwwf2}
\end{eqnarray}

\begin{figure}[htb]
\centering
\epsfig{figure=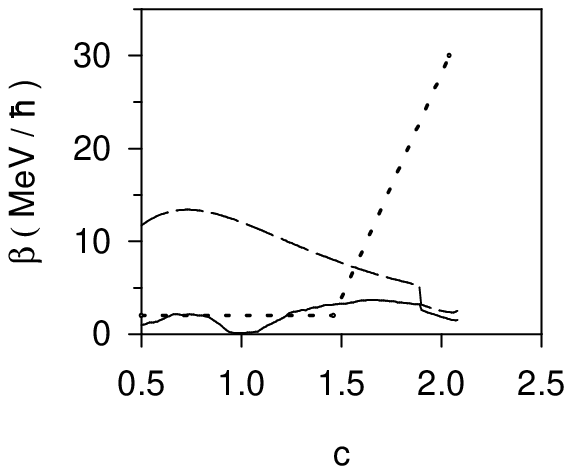}
\caption{\label{fig1}   Reduced   one-body  friction  coefficient
$\beta$ with chaos-weighted wall formula (solid  line)  and  wall
formula    (dashed    line)   frictions   for   $^{213}$Fr.   The
phenomenological reduced coefficient (dotted line) from Ref. [10]
is also shown.}
\end{figure}

Defining  a  quantity  $\beta (c)= \eta (c)/ m(c)$ as the reduced
friction coefficient, its dependence on the elongation coordinate
is shown in Fig.\ref{fig1} for both the WF and CWWF frictions for
the $^{213}$Fr nucleus. A strong suppression of the original wall
formula  friction  for  compact  shapes  of  the  nucleus  can be
immediately noticed in the CWWF friction. This implies  that  the
chaoticity  is very small for near spherical shapes ($c \sim 1$),
the physical picture behind  which  is  as  follows.  A  particle
moving in a spherical mean field represents a typical  integrable
system  and its dynamics is completely regular. When the boundary
of the mean field is set into motion (as in fission), the  energy
gained  by the particle at one instant as a result of a collision
with the moving boundary is eventually fed back to  the  boundary
motion  in  the  course of later collisions. An integrable system
thus becomes completely nondissipative in this picture  resulting
in  a  vanishing  friction  coefficient.  This  aspect  has  been
investigated extensively on earlier occasions \cite{koo,pal1} and
has been found to be valid for any generic integrable system. The
reduction in the strength  of  the  wall  friction  as  shown  in
Fig.\ref{fig1}  thus  becomes  evident from chaos considerations.
The phenomenological reduced friction obtained in Ref.\cite{fro3}
is also shown in this figure. Though the one-body  friction  with
the  CWWF agrees qualitatively with the phenomenological friction
for $c<1.5$, it is beyond its scope to explain the steep increase
of phenomenological friction for $c>1.5$. We shall  discuss  this
point further while presenting the results.

\subsection {Combined dynamical and statistical model calculation}

In our calculation, we first specify the entrance channel through
which  a  compound nucleus is formed. Assuming complete fusion of
the target with the projectile,  the  spin  distribution  of  the
compound nucleus is usually found to follow the analytical form

\begin{eqnarray}
\frac{d \sigma(l)}{dl} = \frac{\pi}{k^2}
\frac{(2l+1)}{1+\exp \frac{(l-l_{c})}{\delta l}}
\label{spin}
\end{eqnarray}

\noindent  where  the parameters $l_{c}$ and $\delta l$ should be
obtained by fitting the experimental fusion cross sections. It is
however found that these parameters for different systems  follow
an  approximate  scaling \cite{fro1} and we shall, therefore, use
the scaled values of these parameters. The initial  spin  of  the
compound  nucleus  will  be  obtained  by sampling the above spin
distribution  function.   The   initial   distribution   of   the
coordinates  and  momenta  $(c,p)$  is  assumed  to  be  close to
equilibrium and  hence  their  initial  values  are  chosen  from
sampling   random   numbers   following   the   Maxwell-Boltzmann
distribution.  With  these  initial  conditions,   the   Langevin
equations  (Eq.\ref{langv})  are numerically integrated following
the procedure outlined in Ref.\cite{ab1}.  The  total  excitation
energy  ($E^{*}$)  of the compound nucleus can easily be obtained
from the beam energy of the projectile and energy conservation in
the form

\begin{equation}
E^{*}=E_{int}+V(c)+p^{2}/2m    \label{temp}
\end{equation}

\noindent gives the intrinsic excitation energy $E_{int}$ and the
corresponding  nuclear  temperature $T=(E_{int}/a)^{1/2}$ at each
time step of integration. The centrifugal potential  is  included
in $V(c)$ in the above equation.

We   shall  also  consider  neutron  and  giant  dipole  $\gamma$
evaporation  at  each  Langevin time step $\tau$ in the following
manner \cite{fro3}. We shall  first  calculate  the  neutron  and
$\gamma$  decay  widths,  $\Gamma_{n}$  and $\Gamma_{\gamma}$, by
using   the   inverse   cross-section   formula   as   given   in
Ref.\cite{fro1}.  These  widths depend upon the temperature, spin
and the mass number of the compound nucleus and hence are  to  be
evaluated  at  each  interval  of  time evolution of the compound
nucleus. We shall next decide whether any evaporation takes place
during the interval or not by first calculating the ratio $x=\tau
/  \tau_{tot}$  where  $\tau_{tot}=  \hbar  /  \Gamma_{tot}$  and
$\Gamma_{tot}=\Gamma_{n}+\Gamma_{\gamma}$. We shall then choose a
random  number  $r$  by sampling from a uniformly distributed set
between 0 and 1. If we find $r < x$, it will  be  interpreted  as
emission  of either a neutron or a $\gamma$ during that interval.
The type of the emitted particle is next decided by a Monte Carlo
selection where it is considered as a neutron if  $0  \le  r  \le
\Gamma_{n}  /  \Gamma_{tot}$,  $r$  being  again  sampled  from a
uniform distribution of random numbers ($0\le r \le 1$), and as a
$\gamma$  otherwise.  This  procedure  simulates   the   law   of
radioactive  decay  for  the emitted particles. The energy of the
emitted particle is then obtained by another Monte Carlo sampling
of its energy spectrum. The intrinsic excitation energy, mass and
spin  of  the  compound  nucleus  are  recalculated  after   each
emission.  The spin of the compound nucleus is reduced only in an
approximate way by assuming  that  each  neutron  or  a  $\gamma$
carries  away  $1\hbar$  angular  momentum. A Langevin trajectory
will be  considered  as  undergone  fission  if  it  reaches  the
scission  point  ($c_{sci}$)  in  course  of  its time evolution.
Alternately it will be counted as an evaporation residue event if
the intrinsic excitation energy becomes smaller than  either  the
fission   barrier  or  the  binding  energy  of  a  neutron.  The
calculation proceeds until the compound nucleus undergoes fission
or ends up as an  evaporation  residue.  The  number  of  emitted
neutrons  and  photons  is  recorded for each fission event. This
calculation  is  repeated  for  a  large   number   of   Langevin
trajectories  and  the  average number of neutrons emitted in the
fission  events  will  give  the  required  prescission   neutron
multiplicity.  The  fission  probability  will be obtained as the
fraction of the trajectories which have undergone fission.

The above scheme can however take an extremely long computer time
particularly  for those compound nuclei whose fission probability
is small. We shall therefore  follow  a  combined  dynamical  and
statistical  model,  first  proposed  by  Mavlitov  {\it  et al.}
\cite{fro2}, in the present calculation. In this model, we  shall
first  follow  the time evolution of a compound nucleus according
to the Langevin equations as described above for  a  sufficiently
long period during which a steady flow across the fission barrier
is  established.  Beyond  this  period,  a  statistical model for
compound nucleus decay is expected to be a equally valid and more
economical in terms of computation.  We  shall  therefore  switch
over to a statistical model description after the fission process
reaches  the  stationary  regime.  We shall, however, require the
fission width along with the neuton and $\gamma$  widths  in  the
statistical  branch of the calculation. This fission width should
be the stationary limit of the fission rate as determined  by the
Langevin equation. Though analytic solutions  for  fission  rates
can  be  obtained  in  special  cases \cite{kram,fro4} assuming a
constant friction, this is not the case with  the  CWWF  friction
which  is  not  constant and is strongly shape dependent. Thus it
becomes necessary to find  a  suitable  parametric  form  of  the
numerically  obtained  stationary  fission  widths using the CWWF
(and also WF) frictions in order to use them in  the  statistical
branch of our calculation. The details of this procedure is given
in  Ref.\cite{pal2}  following  which  we shall calculate all the
required fission widths for the present work.

\section {RESULTS}

\begin{figure}[htb]
\centering
\epsfig{figure=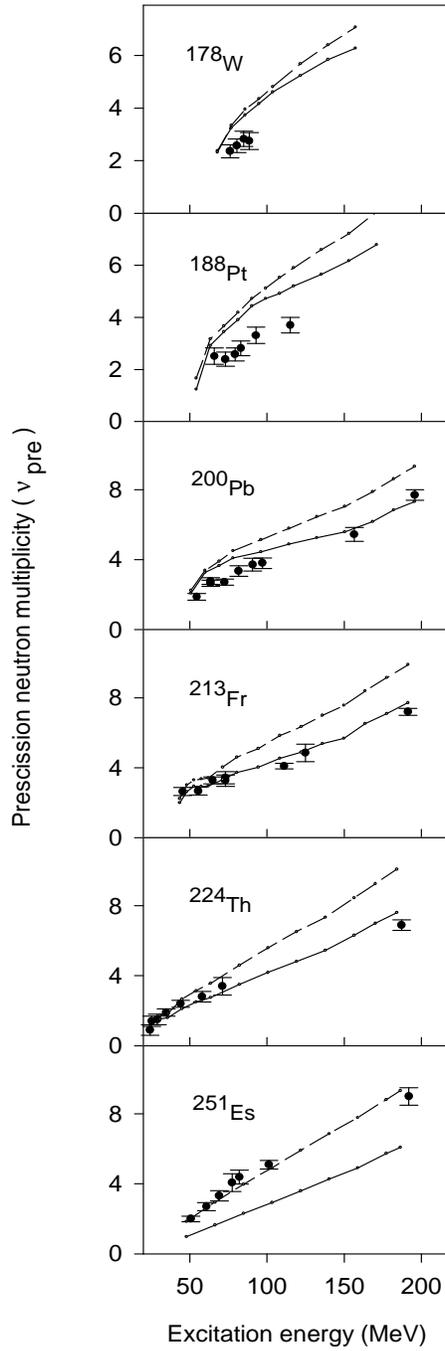,width=6cm,height=18cm}
\caption{\label{fig2}Prescission      neutron      multiplicities
calculated with the CWWF friction are shown as  points  connected
by  solid lines whereas those calculated with the WF friction are
shown as points connected by dashed lines. The experimental  data
for  $^{178}$W,  $^{188}$Pt,  $^{200}$Pb, $^{213}$Fr, $^{224}$Th,
and $^{251}$Es are from Refs. [25],  [25,26],  [25-27],  [25-27],
[26,28], and [25,26],  respectively.}
\end{figure}

 We have  calculated the prescission neutron multiplicity and the
fission probability for a number of  compound  nuclei  formed  in
heavy-ion  induced  fusion  reactions. We have used both the CWWF
and  WF frictions in our calculation. Figure \ref{fig2} shows the
results for  prescission  neutron  multiplicity  along  with  the
experimental  data.  A  number  of  systematic  features  can  be
observed from  these  results.  First,  the  prescission  neutron
multiplicity values calculated with the CWWF and WF frictions are
very  close  at  smaller  excitation  energies,  though at higher
excitation energies,  the  WF  predictions  are larger than those
obtained with the CWWF. This aspect is present in  the  decay  of
all  the  compound  nuclei  which  we  consider  here  and can be
qualitatively understood as follows. The magnitude  of  the  CWWF
friction being smaller than that of the WF friction, fission rate
with  the  CWWF friction is higher than that obtained with the WF
friction. We have shown earlier \cite{pal2} that  the  stationary
fission  width with the CWWF friction is about twice of that with
the WF friction. However at  a  low  excitation  energy  where  a
compound  nucleus is formed with a low value of spin, the fission
barrier is high and both the CWWF and WF fission widths turn  out
to  be  many  times smaller than the neutron width. The neutrons,
therefore, have enough time to be emitted long before a  compound
nucleus  undergoes  fission  irrespective  of  its dynamics being
controlled by either the CWWF  or  the  WF  frictions.  Thus  the
prescission  neutron  multiplicities  are  rather  insensitive to
fission time scales at lower excitation energies.  On  the  other
hand,  a  compound nucleus is formed with a larger spin at higher
excitation energies resulting  in  a  reduction  of  the  fission
barrier.  The fission time scales and the neutron lifetimes start
becoming  comparable  at  higher  excitation  energies  and  less
neutrons are predicted from calculations with the CWWF than those
with  the  WF.  The prescission neutron multiplicity thus becomes
capable of discriminating between  different  models  of  nuclear
friction at higher excitation energies of the compound nucleus.

A  similar explanation also holds for the systematic variation of
the calculated prescission neutron multiplicities with respect to
the mass number of the compound nucleus.  We  find  that  the  WF
prediction  for prescission neutrons starts getting distinct from
that of the CWWF at smaller values of the excitation energy  with
increasing mass number of the compound nucleus. Since the fission
barrier decreases with the increasing mass of a compound nucleus,
the   fission  time  scales  and  the  neutron  lifetimes  become
comparable  for  heavier  compound  nuclei  at  lower  excitation
energies. This results in a fewer neutrons from calculations with
the CWWF than those with the WF as one considers heavier compound
nuclei.

A  number  of interesting points can be noted while comparing the
calculated values  with  the  experimental data. For the compound
nucleus   $^{178}$W,   the    available    experimental    points
\cite{newton}  are  at  low  excitation  energies  and therefore,
cannot distinguish between the calculated values using  the  CWWF
and  WF  frictions,  which  are  almost identical. The calculated
values slightly overestimate the prescission neutron multiplicity
compared to  the  experimental  data.  A  more  extensive  set of
experimental values  for  prescission  neutron  multiplicity  are
available   for   the  compound  nuclei  $^{188}$Pt,  $^{200}$Pb,
$^{213}$F,r and $^{224}$Th \cite{newton,hinde1,hinde2} covering a
wider range of excitation energy in which the  calculated  values
with  the  CWWF and WF differ. Clearly, the CWWF predicted values
give excellent agreement with the  experimental  data  for  these
compound  nuclei  whereas  the  WF  predictions  are considerably
higher. However, similar conclusions  cannot  be  drawn  for  the
heavier  nucleus  $^{251}$Es.  It appears that the WF predictions
are closer to the experimental data \cite{rossner,hinde1,newton}
 whereas the CWWF predictions are somewhat lower. We shall return
to  this  point later for a detailed discussion. For the present,
we  shall   consider   the   results   of   fission   probability
calculations.

\begin{figure}[htb]
\centering
\epsfig{figure=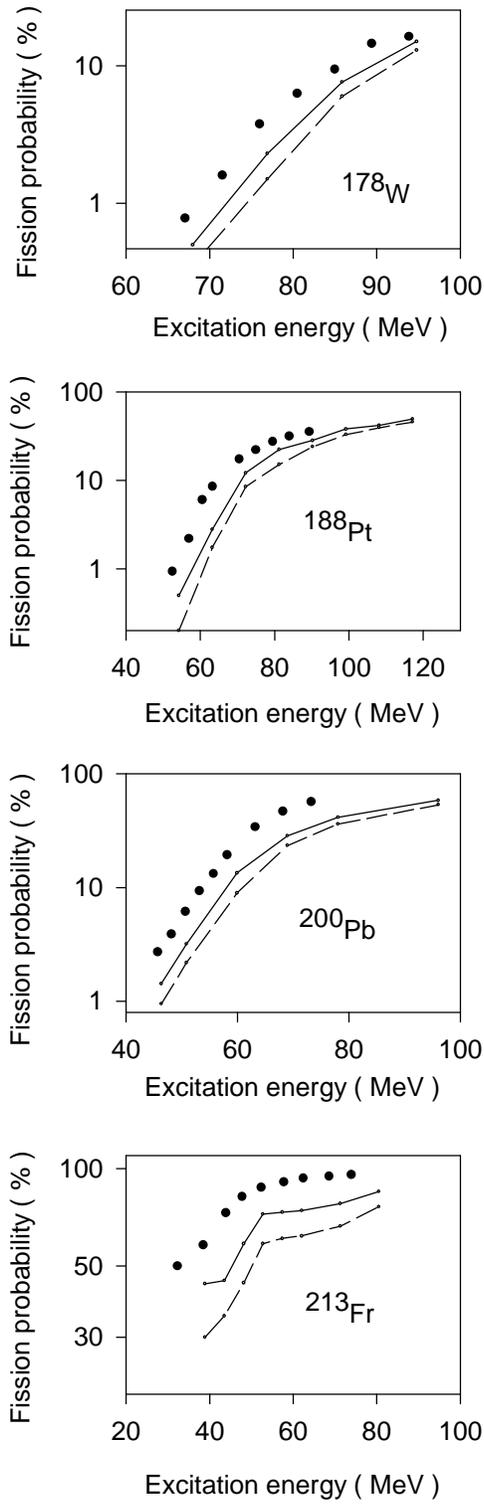}
\caption{\label{fig3}Fission  probabilities  calculated  the CWWF
friction are shown as points connected  by  solid  lines  whereas
those  calculated  with  the  WF  friction  are  shown  as points
connected by dashed lines. The experimental data  for  $^{178}$W,
$^{188}$Pt,$^{200}$Pb,  and $^{213}$Fr are from Refs. [29], [29],
[30], and [31],  respectively.}
\end{figure}

The calculated and experimental values of fission probability are
shown  in  Fig. \ref{fig3} for four compound nuclei. Experimental
data for $^{224}$Th is rather scanty and fission probability  for
$^{251}$Es  is  almost $100 \%$. Hence they are excluded from the
present discussion. The calculated values of fission  probability
complements  the  picture  of fission dynamics which was obtained
while  discussing  the  prescission  neutron  data.  The  fission
probability  is  found  to  be  more  sensitive  to the choice of
friction at lower excitation energies than at higher excitations.
The CWWF predicted fission probabilities are  larger  than  those
from  the  WF  predictions.  Moreover,  the  CWWF predictions are
consistently  closer  to  the  experimental  values  of   fission
probability than those from the WF predictions.

\begin{figure}[htb]
\centering
\epsfig{figure=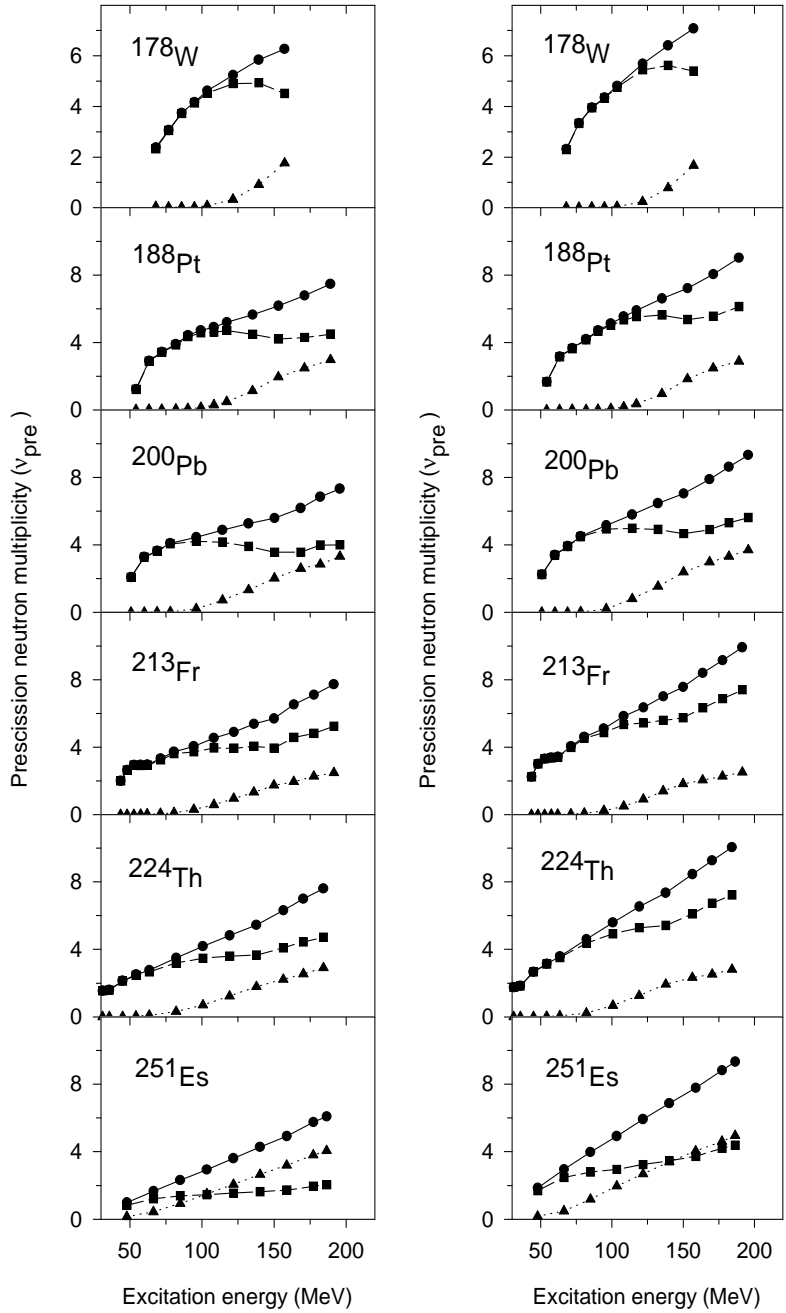}
\caption{\label{fig4}Neutrons  emitted  during  the presaddle and
postsaddle (saddle to scission) stages of fission. Figures in the
left panel show values calculated with the CWWF friction  whereas
those  in  right panel are obtained with the WF friction. In each
plot,  the  solid  circles,  the  solid  squares  and  the  solid
triangles represent the total number of prescission neutrons, the
number  of  presaddle  neutrons  and  the  number  of  postsaddle
neutrons, respectively.}
\end{figure}

In order to gain further insight into the dynamics of fission, we
have  also  calculated  the  presaddle  and postsaddle (saddle to
scission)  contributions  to  the  multiplicity  of   prescission
neutrons.  Figure \ref{fig4} shows the results obtained with both
the CWWF and WF frictions.  For  all  the  cases,  starting  from
almost  zero  multiplicity  at  small  excitation  energies,  the
postsaddle contribution increases at higher excitation  energies.
It is further observed that the postsaddle neutron multiplicities
calculated with the CWWF and WF frictions are almost same for all
the  compound  nuclei  over  the  range  of  excitation  energies
considered here. This would be due to the fact that the number of
postsaddle neutrons depends on the time scale of descent from the
saddle  to  the  scission.  This,  in  turn, will depend upon the
strength of the friction between the saddle and the scission  and
we  have  already  seen  in  Fig. \ref{fig1} that the CWWF and WF
frictions are indeed close at large deformations. We  shall  next
compare  the presaddle contributions calculated with the CWWF and
WF frictions for each  of  the  nuclei  under  consideration.  We
immediately  notice  that  the  WF  predictions  are consistently
larger than those from the CWWF at  higher  excitation  energies.
This gives rise to the enhancement of the WF prediction for total
prescission  multiplicity  compared  to  that   from   the   CWWF
prediction,  which we have already noticed in Fig. \ref{fig2} and
have   discussed   earlier.  Since  the  CWWF  predicted  neutron
multiplicities agree with the experimental values for the  nuclei
$^{178}$W, $^{188}$Pt, $^{200}$Pb, $^{213}$Fr, and $^{224}$Th, we
conclude  that the chaos-weighted wall formula provides the right
kind of friction to describe the presaddle  dynamics  of  nuclear
fission.

\begin{figure}[htb]
\centering
\epsfig{figure=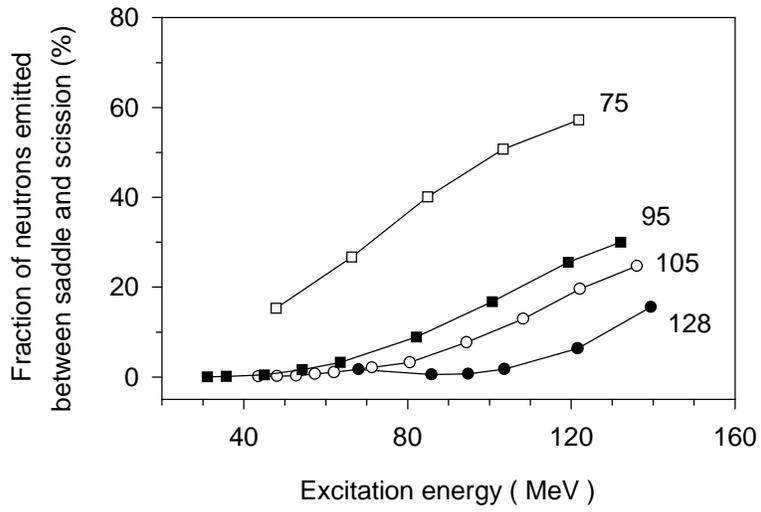}
\caption{\label{fig5} Fraction of neutrons emitted between saddle
and  scission  is  shown  as  a function of excitation energy for
different compound nuclei. The the open square, the solid square,
the open circle and the solid  circle  represent  the  calculated
values  for  $^{251}$Es,  $^{224}$Th,  $^{213}$Fr, and $^{178}$W,
respectively. The critical excitation energy (in units  of  MeV),
as  defined  in  the  text,  is  indicated  for  each  nucleus. }
\end{figure}

While  comparing  the  relative  importance  of the presaddle and
postsaddle neutrons, we further note that the postsaddle neutrons
are more frequently emitted from  heavier  compound  nuclei.  For
$^{251}$Es,  most  of  the  prescission neutrons predicted by the
CWWF are accounted for by the postsaddle neutrons. The underlying
physical picture can be described as  follows.  When  a  compound
nucleus  is  formed  in  a heavy-ion induced fusion reaction, its
spin distribution is assumed to be given by Eq.\ref{spin}. If the
compound nucleus is formed with a  spin  at  which  there  is  no
fission  barrier,  its  transition  to the scission point will be
essentially  considered  as  postsaddle  dynamics.  In  order  to
simplify  our discussion, let us assume that most of the compound
nuclei at a given excitation energy  are  formed  with  the  spin
$l_{0}$  of  Eq.\ref{spin}  and  let $l_{b}$ be the limiting spin
value at which the fission barrier vanishes. We can then  find  a
critical  excitation  energy,  $E_{crit}$,  above  which  $l_{0}$
becomes greater than $l_{b}$ and most of the fission dynamics  at
excitations  above  this  critical  value  can  be  considered as
comprising of only postsaddle trajectories. In  Fig.  \ref{fig5},
we   have  plotted  the  fraction  of  neutrons  emitted  in  the
postsaddle stage as a function of the  excitation  energy  for  a
number  of  compound  nuclei.  The critical excitation energy for
each  nucleus  is  also given in this plot. We have used the CWWF
predicted neutron multiplicities for this plot where we find that
the critical excitation energy decreases  with  increase  in  the
compound  nuclear mass. Thus the dominance of postsaddle neutrons
sets in at lower excitation energies for heavier nuclei which, in
turn, gives rise to the increase in the  fraction  of  postsaddle
neutrons with increasing mass of the compound nucleus.

Though the above discussion clearly establishes the importance of
postsaddle neutrons for a very heavy compound nucleus, the number
of  postsaddle  neutrons  calculated with the CWWF friction still
falls short of making the total prescission multiplicity equal to
the experimental values for $^{251}$Es. We consider the  apparent
better  agreement  between  the  WF predicted prescission neutron
multiplicity and the experimental data for $^{251}$Es as shown in
Fig. \ref{fig2} as a mere coincidence and  we  do  not  find  any
physical  justification  for abandoning the chaos-weighted factor
in one-body friction for such heavy nuclei. Instead, we feel that
the  mechanism  of  neutron  emission  in  the  postsaddle  stage
requires  a  closer  scrutiny  essentially  because  the  nucleus
becomes strongly deformed beyond the saddle  point.  The  neutron
decay  width  of  such a strongly deformed nucleus could be quite
different from that of the  equilibrated  near-spherical  nucleus
which   we   use   in   our   calculation.   In  particular,  the
neutron-to-proton ratio is expected to  be  higher  in  the  neck
region  than  that  in  the  nuclear bulk and this can cause more
neutrons to  be  emitted.  Further,  dynamical  effects  such  as
inclusion  of the neck degree of freedom in the Langevin equation
can influence the time scale of the postsaddle dynamics and hence
the  number  of  emitted  neutrons.  Such possibilities should be
examined in future for a better understanding of  the  postsaddle
dynamics of nuclear fission.

 \section{SUMMARY AND CONCLUSIONS}

We have applied a theoretical model of one-body nuclear friction,
namely   the   chaos-weighted   wall   formula,  to  a  dynamical
description of compound nuclear decay where fission  is  governed
by the Langevin equation coupled with the statistical evaporation
of light particles and photons. We have used both the normal wall
formula  and  its modified form with the chaos-weighted factor in
our calculation in order  to  find  its  effect  on  the  fission
probabilities and prescission neutron multiplicities for a number
of  compound  nuclei.  The  strength  of  the chaos-weighted wall
formula friction  being  much  smaller  than  that  of  the  wall
formula,  the  fission probabilities calculated with the CWWF are
found  to be larger than those predicted with the WF friction. On
the other hand, the  prescission neutron multiplicities predicted
with the CWWF friction turn out to be smaller  than  those  using
the WF friction. Both the prescission  neutron  multiplicity  and
fission  probability  calculated  with  the CWWF friction for the
compound nuclei $^{178}$W,  $^{188}$Pt,  $^{200}$Pb,  $^{213}$Fr,
and  $^{224}$Th  agree  much  better  with  the experimental data
compared to the predictions of the WF friction.

We  have  subsequently  investigated  the  role  of presaddle and
postsaddle  neutrons  at  different   excitation   energies   for
different compound nuclei. It has been shown that the majority of
the  prescission neutrons are emitted in the postsaddle stage for
a very heavy nucleus like $^{251}$Es. The CWWF friction, however,
cannot  produce  enough  neutrons  to  match   the   experimental
prescission  multiplicities for such a nucleus. It is, therefore,
possible that  in  the  postsaddle  region,  either  the  fission
dynamics  gets  considerably slowed down or the neutrons are more
easily emitted. These aspects require further studies  before  we
draw  conclusions  regarding  the  postsaddle dynamics of nuclear
fission.

The presaddle neutrons are however found  to  account for most of
the prescission neutrons for lighter nuclei at  lower  excitation
energies.  On  the  basis  of  the  comparison  of the calculated
prescission multiplicities with experimental data as given in the
preceding section, we can conclude that the  chaos-weighted  wall
formula  friction can adequately describe the fission dynamics in
the presaddle region.

 \section*{ACKNOWLEDGMENTS}
 The   authors are grateful to Nicolas Carjan for making valuable
suggestions during the course of the work.

\eject

\end{document}